\documentclass[conference]{IEEEtran} 
\usepackage[table,xcdraw]{xcolor}
\usepackage{colortbl}
\usepackage{graphicx}
\usepackage{epstopdf}
\usepackage{float}
\usepackage{array}
\usepackage{booktabs} 
\usepackage{amsmath}
\usepackage{amssymb}
\usepackage{mdwmath}
\usepackage{eqparbox}
\usepackage{stfloats}
\usepackage{fixltx2e}
\usepackage{cases}

\usepackage{flushend}

\usepackage{makecell}
\usepackage{algpseudocode}
\usepackage[boxed,ruled,commentsnumbered]{algorithm2e}
\usepackage{url}
\usepackage{cite}
\ifCLASSOPTIONcompsoc
\usepackage[caption=false,font=normalsize,labelfont=sf,textfont=sf]{subfig}
\else
\usepackage[caption=false,font=footnotesize]{subfig}
\fi
\allowdisplaybreaks[4]

\makeatletter
\renewcommand*{\@opargbegintheorem}[3]{\trivlist
      \item[\hskip \labelsep{\bfseries #1\ #2}] \textbf{(#3):}\ }
\makeatother

\begin{document}

\title
{Unleashing the Potential of Beamspace Modulation in Near-Field MIMO: Opportunities and Challenges}

\author{
 \IEEEauthorblockN{Shuaishuai Guo,  Kaiqian Qu}\IEEEauthorblockA{\textit{School of Control Science and Engineering} \\\textit{Shandong University} \\
Email: shuaishuai\_guo@sdu.edu.cn, \\qukaiqian@mail.sdu.edu.cn}
\and
 \IEEEauthorblockN{Shuping Dang} \IEEEauthorblockA{\textit{ School of Electrical, Electronic and}\\\textit{ Mechanical Engineering} \\\textit{University of Bristol}\\
 Email: shuping.dang@bristol.ac.uk}
}

\maketitle


\begin{abstract}
The principal distinction in transitioning from far-field multiple-input multiple-output (MIMO) systems to near-field MIMO configurations lies in the notable augmentation of spatial degrees of freedom (DoF). This increase is not static; rather, it dynamically fluctuates in response to user mobility. A critical challenge emerges in effectively leveraging this significantly enhanced and continuously evolving spatial DoF, particularly when constrained by a limited and energy-intensive array of radio frequency (RF) chains. 
This article presents an exhaustive review of the current methodologies for exploring spatial DoF in MIMO systems, with a particular emphasis on the near-field context. Central to this review is the exploration of beamspace modulation, a technique that ingeniously capitalizes on the increased and dynamic spatial DoFs inherent in near-field MIMO systems. This strategic exploitation is demonstrated to yield significant enhancements in both spectral efficiency and system reliability.
Furthermore, the article delves into a detailed analysis of the multifaceted challenges associated with implementing this technology.  Through this comprehensive evaluation, the work provides crucial insights into ongoing efforts to navigate these challenges and suggests potential pathways for future research in this rapidly evolving field.
\end{abstract}


\section{Introduction} 

\IEEEPARstart{T}{he Sixth} 
 generation (6G) networks are expected to mark a significant technological leap forward, promising to augment spectral efficiency by over ten times in comparison to the fifth generation (5G) networks \cite{dang2020should}. This ambitious enhancement necessitates a substantial increase in the number of antennas integrated into transceivers. Unlike the massive multiple-input and multiple-output (MIMO) communication technology, which is a cornerstone in 5G networks, 6G aspires to implement an even more advanced concept termed \textit{extremely large MIMO} (XL-MIMO) \cite{10098681,Cui2021ChannelEF}. This evolution is critical for facilitating communication in higher-frequency bands, notably the millimeter wave (mmWave) and terahertz (THz) frequency bands \cite{Guo2020}. Such advancements not only represent a paradigm shift in network capabilities but also pose new challenges and opportunities in the research field of wireless communications\cite{10547324}.

The escalation in the number of antennas precipitates not only substantial hardware modifications but also a significant alteration in electromagnetic (EM) radiation patterns \cite{Zhou2015SphericalWC}. EM radiation can be categorized into two distinct regions: the near field and the far field. A pivotal factor in this categorization is the Fraunhofer distance\cite{Selvan2017FraunhoferAF}, which delineates the boundary between these two regions. This distance is directly proportional to the product of the carrier frequency and the square of the array aperture. In the mmWave/THz bands, coupled with the application of XL-MIMO technologies, there is a paradigm shift towards operations predominantly within the near-field region. This signifies a distinction from traditional communication systems that primarily function in the far field\cite{Cui2022NearFieldCF}.

The capability to distinguish propagated signals across varying angles and distances leads to a substantial increase in the number of transmission paths, thereby expanding the spatial degrees of freedom (DoFs) \cite{10098681,10117500}. This enhancement in spatial DoFs, especially when moving from far-field to near-field regions, presents a fascinating area of exploration. A natural initial strategy to capitalize on this increase involves harnessing spatial multiplexing to facilitate the transmission of multiple data streams concurrently \cite{Guo2019}.

Such an approach necessitates the transmitter in the 6G era being outfitted with numerous radio frequency (RF) chains. However, these RF chains are known to be both expensive and power-intensive. In response to these challenges, recent advancements have been made, notably by Wu et al. \cite{Wu2022}, who introduced a distance-aware precoding technique for XL-MIMO systems. This technique signifies a distinction from the traditional hybrid precoding methods \cite{7397861}, which utilize a static number of RF chains, but, instead, dynamically adjust the number of RF chains in operation based on the distance of communications. By doing so, it ensures the maximal exploitation of the available spatial DoFs as the communication distance decreases, thereby significantly improving spectral efficiency while achieving an energy efficiency comparable to those of existing hybrid precoding methods. This innovative precoding strategy selects the optimal beamspace for data transmission, which is directly dependent on the number of RF chains activated. Despite the dynamic nature of such a system, allowing for the selective activation of RF chains, it requires the transceiver to be equipped with an abundant array of RF chains that can be selected, which is difficult to fulfill in practice.

To address this challenge, we delve into the exploration of beamspace modulation (BM) within the near-field region of XL-MIMO systems, aiming to harness the expanded spatial DoFs with only a limited number of RF chains. Unlike the approach proposed by Wu et al., which selects the optimal beamspace corresponding to the number of active RF chains for data transmission, our method introduces non-equiprobable beamspace hopping as a novel technique for conveying additional information. This strategy not only retains the flexibility of dynamically adjusting RF chains but also significantly reduces the required number of RF chains, addressing both cost and energy consumption concerns \cite{GuoWCL}.
This paper presents a comprehensive analysis of BM's effectiveness in the near field region of XL-MIMO, contrasting it with existing methods that solely focus on selecting the best beamspace for transmission. Through evaluations in typical scenarios of near-field communications, we illustrate the superior performance of the BM approach.

\begin{figure*}
    \centering
    \includegraphics[width=0.85\linewidth]{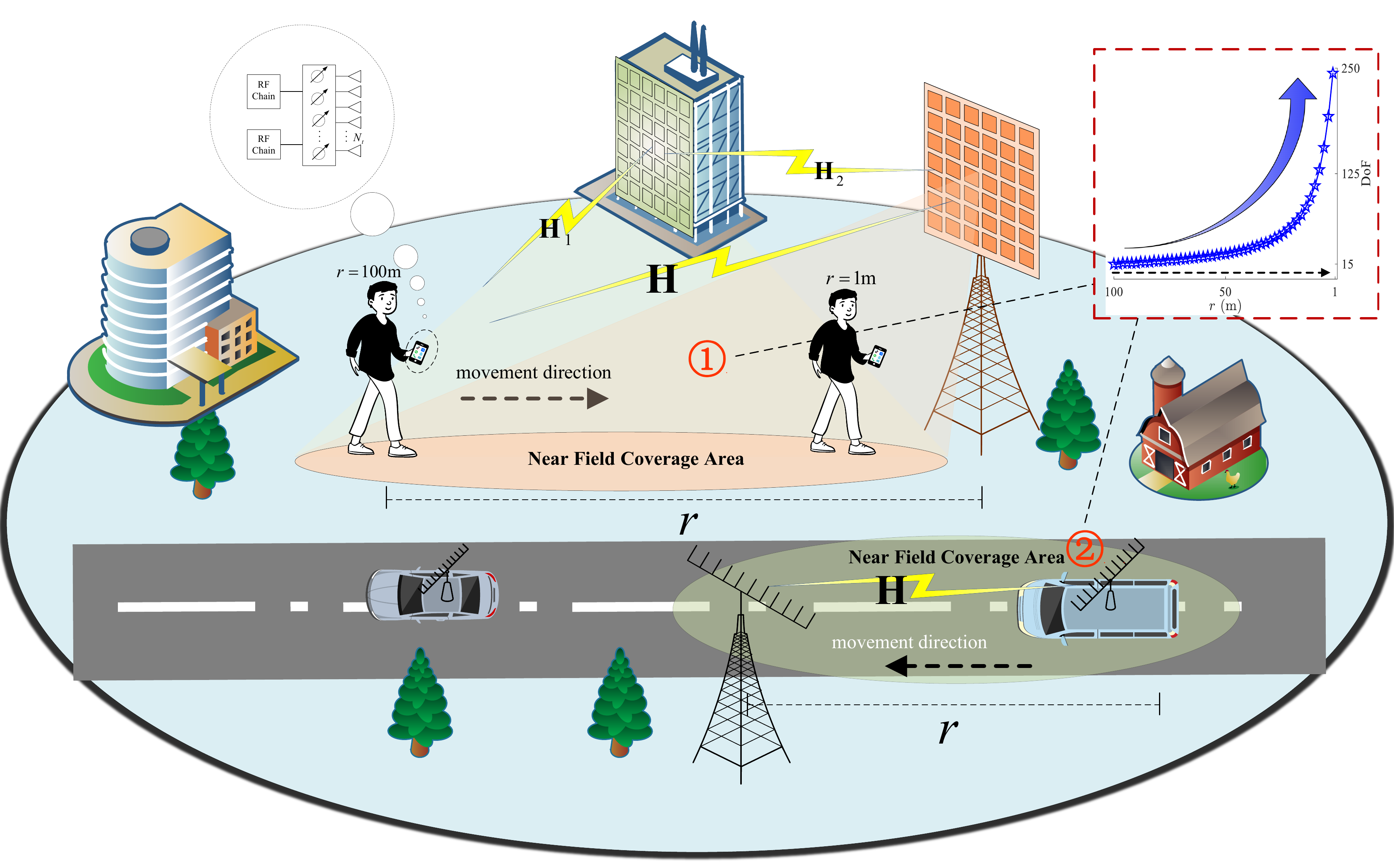}
    \caption{The graphic illustrates an urban wireless network with near-field MIMO communication capabilities. It contrasts the near-field coverage for a pedestrian and a vehicle, both approaching fixed transceiver stations. The accompanying graph details the spatial DoF enhancement from far-field regions to near-field proximity.}
    \label{fig:enter-label}
\end{figure*}

\section{Foundations of Near-Field MIMO and Beamspace Modulation}

\begin{table*}[!t]
\caption{Five Typical Methods for Exploring Spatial DoFs}\label{Tab1}
\renewcommand\arraystretch{1.1}
\begin{tabular}{|c|l|l|l|}
\hline 
\cellcolor[HTML]{6A9DE8}{\color[HTML]{000000} \begin{tabular}[c]{@{}c@{}}~\\ \textbf{DoF Utilization Method} \\ ~ \end{tabular}  } & \multicolumn{1}{c|}{\cellcolor[HTML]{6A9DE8}{\color[HTML]{000000} \begin{tabular}[c]{@{}c@{}}~\\ \textbf{Concept} \\ ~ \end{tabular} }}                                                                                                           & \multicolumn{1}{c|}{\cellcolor[HTML]{6A9DE8}{\color[HTML]{000000} \begin{tabular}[c]{@{}c@{}}~\\ \textbf{Implementation} \\ ~ \end{tabular}}}                                                                                                        & \multicolumn{1}{c|}{\cellcolor[HTML]{6A9DE8}{\color[HTML]{000000} \begin{tabular}[c]{@{}c@{}}~\\ \textbf{Advantage}\\ ~ \end{tabular}}}                                                                     \\ \hline\hline
\rowcolor[HTML]{FFEAAC} 
\cellcolor[HTML]{98C4F1}{\color[HTML]{000000} \textbf{Spatial Multiplexing}}                     & \begin{tabular}[c]{@{}l@{}}Utilizes multiple antennas to transmit \\ different data streams simultaneously\end{tabular}                                                   & \begin{tabular}[c]{@{}l@{}}Common in MIMO systems, transmit\\ different information streams \\ concurrently.\end{tabular}                                                     & \begin{tabular}[c]{@{}l@{}}Increases the data rate without \\ additional bandwidth or transmit \\ power.\end{tabular}                       \\ \hline
\rowcolor[HTML]{FCF0CC} 
\cellcolor[HTML]{98C4F1}{\color[HTML]{000000} \textbf{Spatial Diversity}}                        & \begin{tabular}[c]{@{}l@{}}Employs multiple antennas to transmit \\ the same data over different paths to \\ combat fading and improve signal \\ reliability\end{tabular} & \begin{tabular}[c]{@{}l@{}}Techniques like Alamouti coding \\ are popular, where signals are \\ coded across multiple antennas \\ for diversity gain.\end{tabular}            & \begin{tabular}[c]{@{}l@{}}Enhances signal robustness against \\ interference and fading.\end{tabular}                                      \\ \hline
\rowcolor[HTML]{FFEAAC} 
\cellcolor[HTML]{98C4F1}{\color[HTML]{000000} \textbf{Spatial Modulation}}                       & \begin{tabular}[c]{@{}l@{}}Modulating the signal's spatial \\ characteristics by using the antenna \\ index as an additional dimension of \\ modulation.\end{tabular}     & \begin{tabular}[c]{@{}l@{}}Particularly useful in systems \\ with a large number of antennas, \\ offering a balance between \\ complexity and performance.\end{tabular}       & \begin{tabular}[c]{@{}l@{}}Reduces hardware complexity and \\ consumption by activating only \\ one antenna at a time.\end{tabular}         \\ \hline
\rowcolor[HTML]{FCF0CC} 
\cellcolor[HTML]{98C4F1}{\color[HTML]{000000} \textbf{Beamforming}}                              & \begin{tabular}[c]{@{}l@{}}Manipulating the phase and amplitude \\ of the signal at each antenna to \\ create a directional beam towards \\ the receiver.\end{tabular}    & \begin{tabular}[c]{@{}l@{}}Adaptive beamforming is used \\ cellular networks and Wi-Fi, \\ to dynamically focus the signal \\ where it is needed.\end{tabular}                & \begin{tabular}[c]{@{}l@{}}Improves signal strength at the \\ receiver and mitigates interference \\ to/from other directions.\end{tabular} \\ \hline
\rowcolor[HTML]{FFEAAC} 
\cellcolor[HTML]{98C4F1}{\color[HTML]{000000} \textbf{SDMA}}         & \begin{tabular}[c]{@{}l@{}}Separates users in space to allow \\ multiple users to access the same \\ frequency spectrum simultaneously.\end{tabular}                      & \begin{tabular}[c]{@{}l@{}}Often used in conjunction with \\ beamforming to create distinct \\ beams for different users, thus \\ maximizing spatial separation.\end{tabular} & \begin{tabular}[c]{@{}l@{}}Increases system capacity by serving \\ multiple users concurrently in the \\ same frequency band.\end{tabular}  \\ \hline
\end{tabular}
\end{table*}

 The transition to near-field MIMO is partly driven by the increasing size of antenna arrays, particularly in systems like  XL-MIMO. As these arrays become larger, a significant portion of the communication link may lie in the near-field region. In wireless communications, the region close to the antenna array where the electromagnetic field exhibits rapid spatial variation is known as the near-field region. This contrasts with the far-field region, where the field varies more uniformly and is typically modeled as planar waves. In the near-field region, the wavefronts are spherical rather than planar. This means that the signal's phase varies more dramatically across the antenna array compared to the far-field. The spherical wavefronts in the near-field region allow for finer spatial resolution. This means that signals can be more accurately targeted to specific users or areas, improving spatial multiplexing.

With higher spatial resolution, near-field MIMO can potentially support a greater number of simultaneous data streams. This translates to an increase in system capacity. The precondition is that the transmitter and receiver have enough RF chains to support the simultaneous data stream transmission. The spatial DoFs in the near-field are highly dependent on the user's location relative to the antenna array. The closer the user is to the array, the more pronounced the spherical nature of the wavefront, leading to increased spatial DoFs. In near-field MIMO, the utilization of spatial DoFs becomes more dynamic and adaptable to user locations and movements. This is a shift from far-field MIMO, where spatial DoFs are generally static.

The approximate expression for spatial DoF was derived in \cite{9641865} and can be represented by the following formula:
\begin{equation}
\label{eq1}
    \textrm{DoF}(r)=1+\frac{2L_T}{\lambda \sqrt{1+4\left(\frac{r}{L_R}\right)^2}},
\end{equation}
where $r$ is the distance between the base station (BS) and user, $\lambda$ is the wavelength, $L_T$ is the array apertures of the user, and $L_R$ is the array apertures of the BS.
The validity of this approximation is corroborated by the data presented in Figure 2. This figure illustrates a notable increase in the spatial DoF from $2$ to $70$ as the distance between the transceivers is reduced from $200$ meters to $1$ meter. Particularly noteworthy is the sharp escalation in DoF observed at approximately $40$ meters. To elaborate, the augmentation in DoF from $2$ to $3$, correlating with a decrease in distance from $200$ meters to $40$ meters, is relatively gradual and might even be deemed inconsequential. However, a marked contrast is observed in the range from $40$ meters to $1$ meter, where the DoF surges significantly from $3$ to $70$.
This phenomenon suggests that, in the near-field region, the DoF undergoes a substantial increase only when the distance falls below a certain critical threshold. Utilizing Figure 2 as a reference, this threshold distance can be quantitatively defined using Equation (1) as follows:
\begin{equation}
r_{\Delta}=\frac{L_R}{2}\sqrt{{\left(\frac{2L_T}{(\xi-1)\lambda}\right)^2-1}},
\end{equation}
where $\xi$ is the DoF threshold. In Figure 2, $\xi=3$, it can be obtained that $r_\Delta=40.47 $ m. 

\begin{figure}
    \centering
    \includegraphics[width=1\linewidth]{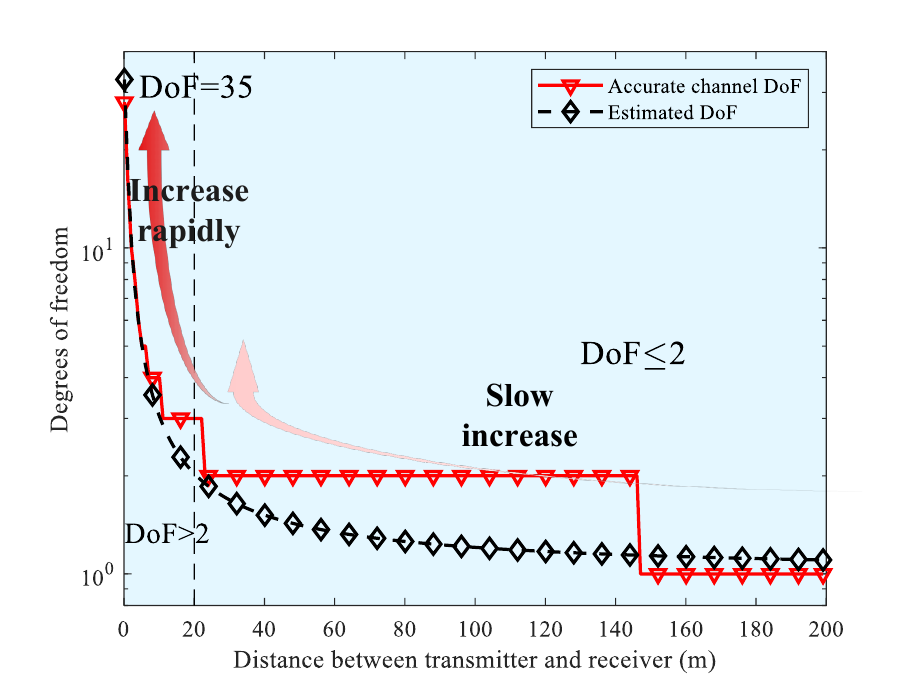}
    \caption{A visual representation of the DoF variation with the changing distance between the transmitter and receiver. It showcases two distinct regions: a moderate increase in DoF at greater distances and a significant upsurge as the system enters the near-field zone, marked by distances less than 40 meters. In the simulation, the user is equipped with a ULA of $128$ antennas, while the BS features a ULA of $256$ antennas spaced half a wavelength apart, operating at $28$ GHz.}
    \label{fig2}
\end{figure}
 
The exploration of significantly enhanced and dynamically evolving spatial DoFs in near-field MIMO systems is paramount for augmenting spectral efficiency and system dependability. Over time, five predominant methodologies utilizing spatial DoFs in distinctive manners have been developed: spatial multiplexing, spatial diversity, spatial modulation, beamforming, and SDMA. Refer to Table \ref{Tab1} for a comprehensive overview.

In addition to these primary methods, various transformations and amalgamations exist. For instance, generalized spatial modulation, which activates multiple antennas while employing antenna group indices for modulation, can be integrated with either spatial multiplexing techniques \cite{GuoGenSM} or spatial diversity approaches \cite{GuoGSM}. A widely recognized strategy in point-to-point near-field MIMO, BBS, traditionally deemed 'optimal', essentially merges beamforming with spatial multiplexing. This approach selectively employs the best $K$ spatial DoFs from a total of $N$ DoFs for spatial multiplexing, where $K$ denotes the count of RF chains. However, our prior research \cite{Guo2019, Guo2020} theoretically and empirically contests the optimality of BBS, suggesting that BM is superior in scenarios with limited transmitting RF chains.

BM amalgamates the concepts of beamforming, spatial modulation, and spatial multiplexing. Contrary to the conventional approach of selecting only $K$ DoFs from $N$ for multiplexing—thereby leaving $N-K$ DoFs underutilized—BM exploits all $\left(N \atop K\right)$ combinations of DoFs. BM uniquely conveys additional information by encoding data onto DoF group indices. Given the varied channel quality across spatial DoF groups, BM strategically activates more efficient DoF groups with a higher probability and less efficient ones with a lower probability. The optimal DoF group index activation probabilities, which we have determined to maximize capacity, were presented in \cite{GuoWCL}.

BM's approach markedly contrasts with that of traditional spatial modulation, which relies on antenna indices for modulation and activates only a limited number of antennas at any given coherent time, thereby not compromising the beamforming gain of antenna arrays. Furthermore, unlike other existing beamformer/precoder index-based modulations that indiscriminately activate both effective and less effective beamspace, BM significantly enhances spectral efficiency by discriminating between these spaces.

\section{Opportunities Afforded by Beamspace Modulation in Near-Field MIMO}

The integration of  BM into near-field MIMO systems marks a notable advancement in system performance, especially in terms of data throughput and SER. The use of BM, leveraging augmented spatial DoF, allows for the simultaneous transmission of larger data volumes without necessitating additional RF chains. This efficiency in multiplexing not only enhances the overall data rate, crucial for high-bandwidth applications, but also maintains a consistent number of RF chains, facilitating a versatile RF configuration suitable for both near-field and far-field communication ranges.

In a typical setup, the user equipment is equipped with a ULA of $48$ antennas and one RF chain, while the BS features a ULA of $256$ antennas spaced half a wavelength apart, operating at $28$ GHz. In such a scenario, with a SNR set at $20$ dB and $\xi=2$, the threshold distance $r_\Delta$ is computed as 33.46 meters.

Figure 3 presents an interesting observation where the spectral efficiency of both BM and BBS schemes increases as the distance decreases, factoring in large-scale fading effects. The BM scheme consistently outperforms the BBS in spectral efficiency, particularly notable as the distance shortens from $50$ meters to $1$ meter, with the efficiency gain of BM rising from $2$\% to $20$\%. This increase is attributed to BM's ability to utilize a greater number of spatial DoFs at shorter distances, leading to more significant gains. However, beyond the threshold distance of $r_\Delta$, the limited spatial DoFs result in only marginal gains from BM.
\begin{figure}
    \centering
    \includegraphics[width=1 \linewidth]{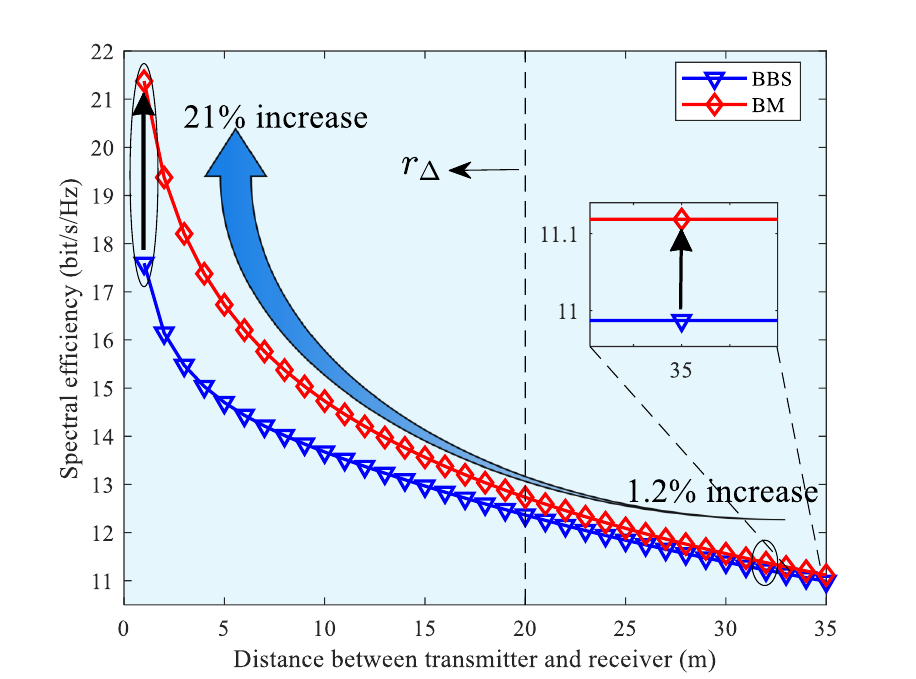}
    \caption{A simulated analysis of spectral efficiency, depicting its dynamic response to changing distances between the transmitting and receiving antennas, thereby offering insights into the efficiency of signal transmission over varied spatial separations}
    \label{fig3}
\end{figure}

\begin{figure}
    \centering
    \includegraphics[width=1 \linewidth]{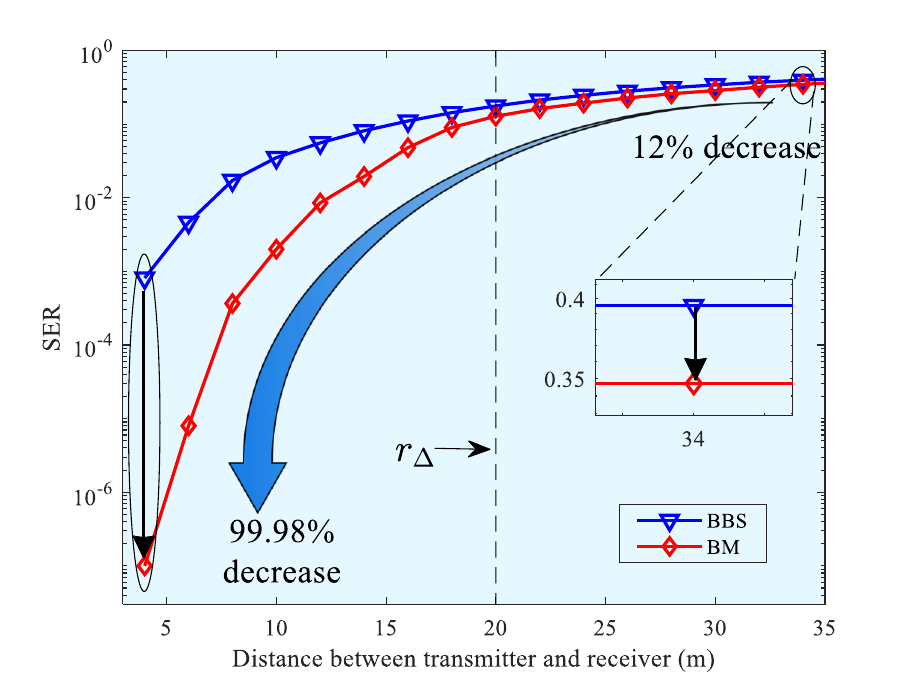}
    \caption{A simulated view of the correlation between the distance separating transceivers and the resulting SER, thereby illustrating the critical impact of physical distance on communication system reliability.}
    \label{fig4}
\end{figure}

Furthermore, BM's advantage is also evident in SER reduction. Figure 4 shows that the SER for both BM and BBS schemes decreases with the reduction in distance between the transmitter and receiver. Remarkably, at a distance of 5 meters, BM's SER is reduced by two orders of magnitude compared to that of the BBS scheme. This aligns with the spectral efficiency trends observed in Figure 3, where shorter distances enable BM to achieve greater reductions in SER, owing to its efficient utilization of spatial DoFs.

Overall, BM's ability to effectively separate signals representing different bits, utilizing the additional spatial DoFs, leads to an enhanced decoding process, greatly improving the decoding SER. This makes BM an advantageous approach in near-field MIMO systems, especially for applications requiring high data throughput and reliability.

\section{Challenges in Implementing Beamspace Modulation in Near-Field MIMO}
The integration of BM within near-field MIMO systems offers a promising avenue for enhancing communication capabilities. However, it also brings to the fore complex challenges that span the domains of hardware capability, algorithmic design, RF chain optimization, and security. 

\subsection{Fast Beam Switching}

The hardware requisite for BM is formidable, particularly because of the necessity for rapid beam switching. Unlike traditional beamforming, which can afford the luxury of switching beams at relatively leisurely coherent times, BM operates on a much tighter time scale, necessitating beam switches at symbol times. These durations can be incredibly brief, measured in microseconds (µs) or even nanoseconds (ns), pushing the limits of current hardware capabilities. High-speed beam switching imposes rigorous demands on computational capacity and the sophistication of algorithms. Real-time calculations and adjustments of beamforming weights must be executed flawlessly to synchronize with the rapid pace of communication signals.

\subsection{Sufficient Receiving RF Chains}

The potential of BM in near-field MIMO is best realized with an adequate number of receiving RF chains. Although BM liberates the system from the necessity of matching transmitting RF chains to the spatial DoFs, the receiving end still requires a substantial number of chains to harness the full capabilities of the increased spatial DoFs.
Receiving RF chains, being less expensive and less power-intensive than their transmitting counterparts, present a more manageable challenge. A near-field MIMO system can still reap the benefits of BM, even with fewer RF chains than spatial DoFs, provided the number of receiving RF chains surpasses the transmitting ones. As depicted in Figure 5, the BM strategy yields performance enhancements as long as the count of RF chains at the receiver exceeds one, which is the number of transmitting RF chains. The performance of the BM approach reaches a saturation point when the count of RF receiving chains aligns with the spatial DoF. This occurs because the active spatial DoFs are effectively limited by the number of receiving chains. Consequently, this calls for a deliberate strategy in RF chain allocation, with an emphasis on prioritizing essential communication links.

\begin{figure}
    \centering
    \includegraphics[width=1\linewidth]{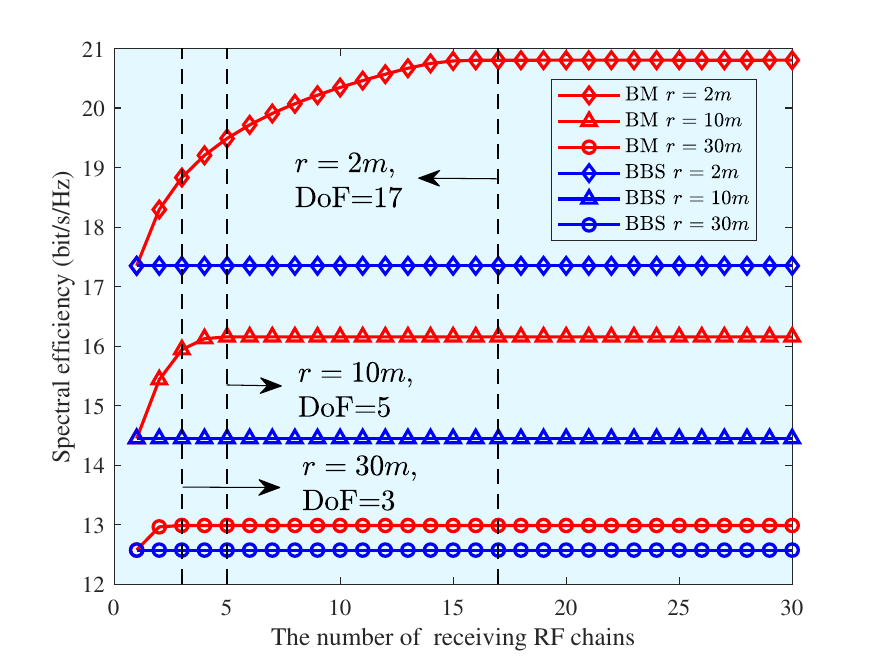}
    \caption{Effect of the number of RF chains at the receiver on spectral efficiency. The relevant setup is the same as the experiment in Figure 3.}
    \label{Effect of the number of RF chains at the receiver on spectral efficiency.}
\end{figure}
\subsection{Interference Leakage and Security Concern}

BM's very mechanism of encoding information onto beam hopping sequences introduces unique security vulnerabilities. Rapid beam hopping can result in periods where the beam's focus is comparatively lax, resulting in signal spillage beyond the intended coverage area. This spillage is not merely a waste of signal; it poses a significant security risk, as it could be intercepted or analyzed to uncover sensitive information about the communication's nature and participants.
Moreover, the pattern of beam activations, if not sufficiently secured, could be leveraged by eavesdroppers to deduce user information. Addressing these security concerns involves implementing robust encryption protocols, designing sophisticated beam hopping patterns that are less susceptible to interception, and developing signal masking techniques to obfuscate the transmission characteristics.

\section{Future Research Directions}
This section delineates potential research directions arising from the challenges and trends identified in the realm of BM within near-field MIMO systems.

\subsection{Multi-User Beamspace Modulation}
The implementation of BM in point-to-point near-field MIMO systems has demonstrably augmented spectral efficiency and reinforced transmission reliability. Extending these benefits to multi-user environments, while appealing, confronts significant challenges, predominantly from various forms of interference such as co-channel and inter-user interference. Addressing these challenges necessitates comprehensive research focused on developing sophisticated interference mitigation strategies, which are crucial for the optimal functioning of these systems.

Key to this endeavor is the strategic allocation of spatial resources among users, leveraging the increased spatial DoFs inherent in BM. A promising strategy involves classifying users based on the orthogonality of their channel state information (CSI), enabling the efficient grouping of users with orthogonal channel vectors to minimize interference and maximize spatial DoFs utilization. This approach should be balanced with considerations of quality of service (QoS), prioritizing users with more stringent data rate and latency requirements, while ensuring fairness in resource distribution. 

\subsection{Distance-aware Beamspace modulation}
In the near-field, the DoFs vary with the distance between the transmitter and receiver. The DoFs and the number of RF chains together determine the beamformer candidate set, thereby affecting the performance of BM. To better leverage BM in dynamically changing distance environments, distance-aware beamspace modulation (DABM) is considered. It is necessary to establish the relationship between BM performance and the number of RF chains, and to find the number of RF chains that maximizes BM gain. By introducing switch circuits, the optimal number of RF chains can be activated based on different distance-related DoFs.

 \subsection{Compressed Sensing-based Detection Methods}
In near-field MIMO systems, BM, while enhancing data rates, concurrently elevates the complexity of detection. For applications demanding real-time or near-real-time communication, such as autonomous vehicles, telemedicine, and real-time video streaming, the development of simplified detection methods is critical. BM’s inherent signal sparsity—characterized by the activation of only a subset of beams at any given time—presents a unique opportunity for the application of compressed sensing (CS). Research in this area should focus on the mathematical modeling of BM signal structures and the CS reconstruction process, tailoring existing CS algorithms to the unique characteristics of BM signals, and empirically validating these algorithms in laboratory settings using hardware that simulates real-world BM systems.
\subsection{Codebook-based Beamspace Modulation}
BM necessitates computationally intensive procedures like the singular vector decomposition of the CSI matrix. Investigating codebook-based BM is thus imperative to reduce computational demands, particularly in XL-MIMO systems. This approach involves utilizing predefined sets of beam patterns, simplifying implementation in practical systems. In XL-MIMO environments, where the number of antenna elements is substantial, codebook-based beamforming can efficiently manage beam directionality without intricate per-element computations. The research should encompass the creation of optimal beam pattern codebooks, evaluating their impact on system performance metrics such as throughput, coverage, and QoS, and ensuring adaptability to diverse environments. Additionally, the feasibility of hardware implementation, considering factors like processing power and energy efficiency, is crucial. The development of algorithms for optimal beam selection from the codebook, based on user location, channel conditions, and system load, is vital. Ensuring compatibility with current and emerging wireless communication protocols and standards is also essential for the broader adoption and integration of these technologies.

\section{Conclusion}
In this paper, we have presented BM as an innovative solution to address the challenges inherent in near-field MIMO systems, particularly the management of significantly increased and dynamically varying spatial DoFs within the constraints of a fixed number of RF chains. Our investigation has substantiated the efficacy of BM, demonstrating its superior capability in augmenting capacity and enhancing the SER when compared to existing benchmarks.
Furthermore, we have thoroughly examined the practical challenges associated with the implementation of BM in near-field MIMO environments. Moreover, we delineate several critical avenues for future research. These include developing strategies to mitigate the implementation challenges and exploring the potential synergies between BM and other emergent technologies. This exploration is essential for advancing the field of near-field MIMO and fully realizing the potential of BM in next-generation wireless communication systems.

\bibliographystyle{IEEEtran} 
\bibliography{IEEEabrv,bib}
\end{document}